\documentclass[a4paper]{jpconf}
\usepackage{graphicx}
\usepackage{epsfig}
\usepackage{epstopdf}
\usepackage{caption}
\usepackage{subcaption}
\usepackage{array}
\usepackage{siunitx}
\usepackage[usenames, dvipsnames]{color}
\usepackage{color}

\usepackage{longtable}
\usepackage[utf8]{luainputenc}
\usepackage{amsmath}
\usepackage{xspace}
\usepackage{multicol}

\usepackage{makecell}
\usepackage{cite}

\newcommand{\Smilei}{{\sc Smilei}\xspace}
\begin{document}
\title{ Azimuthal decomposition study of a realistic laser profile for efficient modeling of Laser WakeField Acceleration}

\author{ I. Zemzemi$^1$, F. Massimo$^1$, A. Beck$^1$}

\address{$^1$ Laboratoire Leprince-Ringuet – École polytechnique, CNRS-IN2P3, Palaiseau 91128, France}

\ead{izemzemi@llr.in2p3.fr}

\begin{abstract}
The advent of ultra short high intensity lasers has paved the way to new and promising, yet challenging, areas of research in the laser-plasma interaction physics. The success of constructing petawatt femtosecond lasers, for instance the Apollon laser  in France, will help understanding and designing future particle accelerators and next generation of light sources. Achieving this goal intrinsically relies on the combination between experiments and massively parallel simulations. So far, Particle-In-Cell (PIC) codes have been the ultimate tool to accurately describe the laser-plasma interaction especially in the field of Laser WakeField Acceleration (LWFA) . Nevertheless, the numerical modelling of laser plasma accelerators in 3D can be a very challenging task. This is due to the large dispersity between the scales involved in this process. 
%This kind of simulations is computationally very expensive: it requires thousands of nodes on super computers.
 In order to make such simulations feasible with a significant speed up, we need to use reduced numerical models which simplify the problem while retaining a high fidelity. Among these models, Fourier field decomposition in azimuthal modes for the cylindrical geometry  \cite{LIFSCHITZ20091803}  is a promising reduced model especially for physical problems that have close to cylindrical symmetry which is the case in LWFA.
This geometry has been implemented in the open-source code \Smilei \cite{DEROUILLAT2018351} in Finite Difference Time Domain (FDTD) discretization scheme for the Maxwell solver. 
%In order to investigate the accuracy of this method, we study the impact of different physical and numerical parameters involved in the simulation done using this geometry. The obtained results are benchmarked with a full 3D cartesian description with high resolution. 
%This study is unique in terms of using data to describe the laser in a realistic way with all its defaults instead of using a gaussian analytical expression. The data has been extracted by our collaborators from the experimental group of Apollon laser.
In this paper we will study the case of a realistic laser measurement from Apollon facility, the ability of this method to describe it correctly and  the determination of the  necessary number of modes for this purpose. We will also show the importance  of higher modes inclusion in the case of realistic laser profiles to insure fidelity in simulation.
\end{abstract}

\section{Introduction}

The continuous upgrade in laser power has permitted the investigation and the verification of new methods of particle acceleration by taking advantage from the high gradient wakefield created when the laser propagates trough an under-dense plasma. Laser WakeField Acceleration (LWFA)\cite{TajimaDawson79}\cite{Esarey2009} \cite{Malka2002} has been proven to be a promising efficient way to accelerate electrons  up to few Gevs %However, the field of laser plasma interaction is very wide and researchers are aiming to reach the 10 Gev%
within a short propagation of distance with high quality of beam. In order to investigate the different experimental  set-ups and to determine the optimal parameters to achieve this goal, simulation is the key to perform a parametric scan and  analyze regimes that haven't been explored yet.
That one may model correctly  the interaction between the laser and the plasma, a full kinetic description of the plasma is needed. Particle-in-Cell code is  ubiquitously used as an established tool that solves Vlasov equation for the different species presented in the plasma coupled with Maxwell equations. It is a powerful method  that gives an accurate description of the plasma response to the laser and captures a wide range of physical phenomena \cite{BirdsallLangdon2004}. Nevertheless,  precise and realistic results  are obtained only with full 3D description with high resolution.  Even though 2D simulations are used in the context of 2D Cartesian slab or in the cylindrical geometry r-z to illuminate the physics, there is a qualitative and quantitative difference with the 3D simulation especially in the case of LWFA  when studying non linear regime. This is mainly because self focus and  self modulation of the phase are not well described by a 2D simulation \cite{Davoine2008}.  In spite of the necessity of using  well resolved 3D simulation  for the accurate description, using many points in the grid for long propagation distance with small time steps is very costly and pushes the existent computing resources available nowadays to the limits in order to have the simulation results in a reasonable   amount of time. Therefore, there have been many methods  suggested  to overcome  this obstacle among which we mention  moving window, quasi-static approximation \cite{Mora1997}, ponderomative guiding center or envelope  description \cite{Benedetti2010}, boosted frame \cite{Vay2007} ... Each method has its advantages and its limits depending on the case of study. 

Thanks to the close to cylindrical symmetry of the laser in LWFA , an alternative has been proposed to reduce the cost of  the simulations while retaining high fidelity \cite{LIFSCHITZ20091803}. In section 2,  the main features of the algorithm are recalled and section 3  shows a study of the number of modes required to mimic the realistic  Apollon laser  profile correctly.

\section{Review of the Azimuthal Fourier decomposition in cylindrical geometry}

 In this algorithm, the fields $\textbf{E}, \textbf{B}, \textbf{J}, \rho$  in cylindrical coordinates $(r,z , \theta )$ are decomposed in Fourier space in $\theta$ direction according to  \eqref{eq:chap2:azimuthal} and \eqref{fourier_decomposition}. 

\begin{equation}
F(r,z, \theta) = \sum_{m=-\infty}^\infty
  \tilde{F}_{m}(r,z) e^{-im\theta} 
\qquad  \textrm{with} 
\qquad \tilde{F}_{m}(r,z) = \frac{1}{2 \pi} \int_0^{2\pi} d\theta
\,F(r,z, \theta)e^{im\theta} 
%\label{eq:chap2:Fourier-coeffs}
\label{eq:chap2:azimuthal}
\end{equation}

\begin{align*}
F(r,z, \theta) &=  \tilde{F}_{0}(r,z) + \sum_{m=1}^\infty \mathrm{Re}\left[ 2 
  \tilde{F}_{m}(r,z) e^{-i m\theta} \right] \\
    &=  \tilde{F}_{0}(r,z) + \sum_{m=1}^\infty \mathrm{Re}\left[ 2 
  \tilde{F}_{m}(r,z) \right]  cos(m \theta) + \mathrm{Im}\left[ 2
  \tilde{F}_{m}(r,z) \right] sin(m \theta) \tag{2}
\label{fourier_decomposition}
\end{align*}

This expansion is replaced in Maxwell equations and thanks to their linearity, it  generates a set of equations \eqref{Maxwell}.   

 \begin{align*}
\frac{\partial \tilde{B}_{r,m} }{\partial t} &=
\frac{im}{r}\tilde{E}_{z,m} + \frac{\partial
  \tilde{E}_{\theta,m}}{\partial z} \\[3mm]
\frac{\partial \tilde{B}_{\theta,m} }{\partial t} &=
 - \frac{\partial \tilde{E}_{r,m}}{\partial z} + \frac{\partial
  \tilde{E}_{z,m}}{\partial r} \\[3mm]
\frac{\partial \tilde{B}_{z,m} }{\partial t} &=
- \frac{1}{r} \frac{\partial (r\tilde{E}_{\theta,m})}{\partial r} - \frac{im}{r}\tilde{E}_{r,m} \\[3mm]
\frac{1}{c^2} \frac{\partial \tilde{E}_{r,m} }{\partial t} &=
-\frac{im}{r}\tilde{B}_{z,m} - \frac{\partial
  \tilde{B}_{\theta,m}}{\partial z} - \mu_0 \tilde{J}_{r,m} \\[3mm]
\frac{1}{c^2}\frac{\partial \tilde{E}_{\theta,m} }{\partial t} &=
 \frac{\partial \tilde{B}_{r,m}}{\partial z} - \frac{\partial
  \tilde{B}_{z,m}}{\partial r} - \mu_0 \tilde{J}_{\theta,m} \\[3mm]
\frac{1}{c^2}\frac{\partial \tilde{E}_{z,m} }{\partial t} &=
 \frac{1}{r} \frac{\partial (r\tilde{B}_{\theta,m})}{\partial r} +
 \frac{im}{r}\tilde{B}_{r,m} - \mu_0 \tilde{J}_{z,m} \tag{3}
 \label{Maxwell} 
\end{align*}

Each mode evolves independently in vacuum and different modes are coupled only when the plasma is present. The Fourier series in equations \ref{eq:chap2:azimuthal} and \ref{fourier_decomposition} is usually  truncated up to very first modes in the case of low dependence on $\theta$.  For example,  wakefields can be described by mode 0  because it is independent from $\theta$ and the laser by mode 1.  In fact, for a cylindrically symmetric pulse (for example a gaussian one) propagating in $z$  and polarized linearly along $\vec{e_{\alpha}} = cos(\alpha)\vec{e_x} + sin(\alpha)\vec{e_y}$ , the field component $\vec{E}$ depends on $\theta $ according to the following description:

\begin{align*}
\vec{E} &= E_0(r,z)\vec{e}_\alpha \\
& = E_0(r,z) [\; \cos(\alpha)(\cos(\theta)\vec{e}_r -
\sin(\theta)\vec{e}_\theta) \;+\; \sin(\alpha)(\sin(\theta)\vec{e}_r +
\cos(\theta)\vec{e}_\theta) \; ]\\
& = \mathrm{Re}[ \; E_0(r,z) e^{i\alpha} e^{-i\theta} \; ]\vec{e}_r \;
+ \; \mathrm{Re}[ \; -i E_0(r,z) e^{i\alpha} e^{-i\theta} \; ]\vec{e}_\theta \\
& = \; E_r \vec{e_r} \;+\; E_{\theta} \vec{e_{\theta}} \;
 \tag{4}
\end{align*}

Here the amplitude $E_0$ does not depend on $\theta$ because the pulse was assumed to be cylindrically symmetric. In this case, the above relation shows that the fields $E_r$ and $E_\theta$ of the laser are represented
exclusively by the mode $m = 1$ and the same stands  for $B_r$ and $B_\theta$. Thus, any cylindrically symmetric laser can be thoroughly  described by the mode 1. As a consequence, the infinite sum of modes can be truncated at the first two modes since  only the modes
$m = 0$ and $m = 1$ are necessary to model laser-wakefield acceleration with linearly polarized  lasers with axisymmetric envelope  such as gaussian ones.

However, realistic lasers are not perfectly gaussian and because of their imperfections and spatio-temporal couplings \cite{Pariente:15} \cite{Jeandet:18} they lose their symmetry and thus an accurate model of realistic laser profiles includes more than just two modes.
This is why, $m_{max}$ the maximal number of modes used  is kept as a free parameter in the implementation of the method.

\section{Experimental laser focal spot data analysis } 
In this section, the optimal number of modes to include in simulations is defined and the impact of the laser envelope imperfections on the decomposition is shown. For this purpose, the experimental intensity distribution in the focal plan from Apollon laser is  used as input in this study. However, spatio-temporal distortions, i.e  spatial dependencies of the temporal properties are not taken into consideration. A gaussian time envelope  was assumed in this study.
The figure  \ref{laser} shows the measured intensity  in the focal plan of the Apollon laser. Intensity is normalized to laser strength parameter $a_0$ . The raw data from the camera was interpolated to the simulation grid using a quadratic interpolation function taking into account the number of pixels of the camera and the pixel size. 
 Table \ref{table}  sums up the laser parameters for this analysis.

%\begin{figure}[h!]
%  \centering
%    \begin{subfigure}[b]{0.5\linewidth}
%    \includegraphics[width=\linewidth]{Bz_laser_center_filtered.png}
%    \caption{Too much coffee.}
%  \end{subfigure}
% \\
%  \begin{subfigure}[b]{0.3\linewidth}
%    \includegraphics[width=\linewidth]{Reconstructed_Bz00.eps}
%     \caption{Coffee.}
%  \end{subfigure}
%  \begin{subfigure}[b]{0.3\linewidth}
%    \includegraphics[width=\linewidth]{Reconstructed_Bz01.png}
%    \caption{More coffee.}
%  \end{subfigure}
%  \begin{subfigure}[b]{0.3\linewidth}
%    \includegraphics[width=\linewidth]{Reconstructed_Bz03.png}
%    \caption{Tasty coffee.}
%  \end{subfigure}
%  
%   \begin{subfigure}[b]{0.3\linewidth}
%    \includegraphics[width=\linewidth]{Reconstructed_Bz04.png}
%     \caption{Coffee.}
%  \end{subfigure}
%  \begin{subfigure}[b]{0.3\linewidth}
%    \includegraphics[width=\linewidth]{Reconstructed_Bz07.png}
%    \caption{More coffee.}
%  \end{subfigure}
%  \begin{subfigure}[b]{0.3\linewidth}
%    \includegraphics[width=\linewidth]{Reconstructed_Bz09.png}
%    \caption{Tasty coffee.}
%  \end{subfigure}
%  \caption{The same cup of coffee. Multiple times.}
%  \label{fig:coffee3}
%\end{figure}
\begin{table}
\begin{minipage}{.45\textwidth}
%\begin{figure}[h!]
\centering
\captionsetup{justification=centering}
\includegraphics[width=\linewidth]{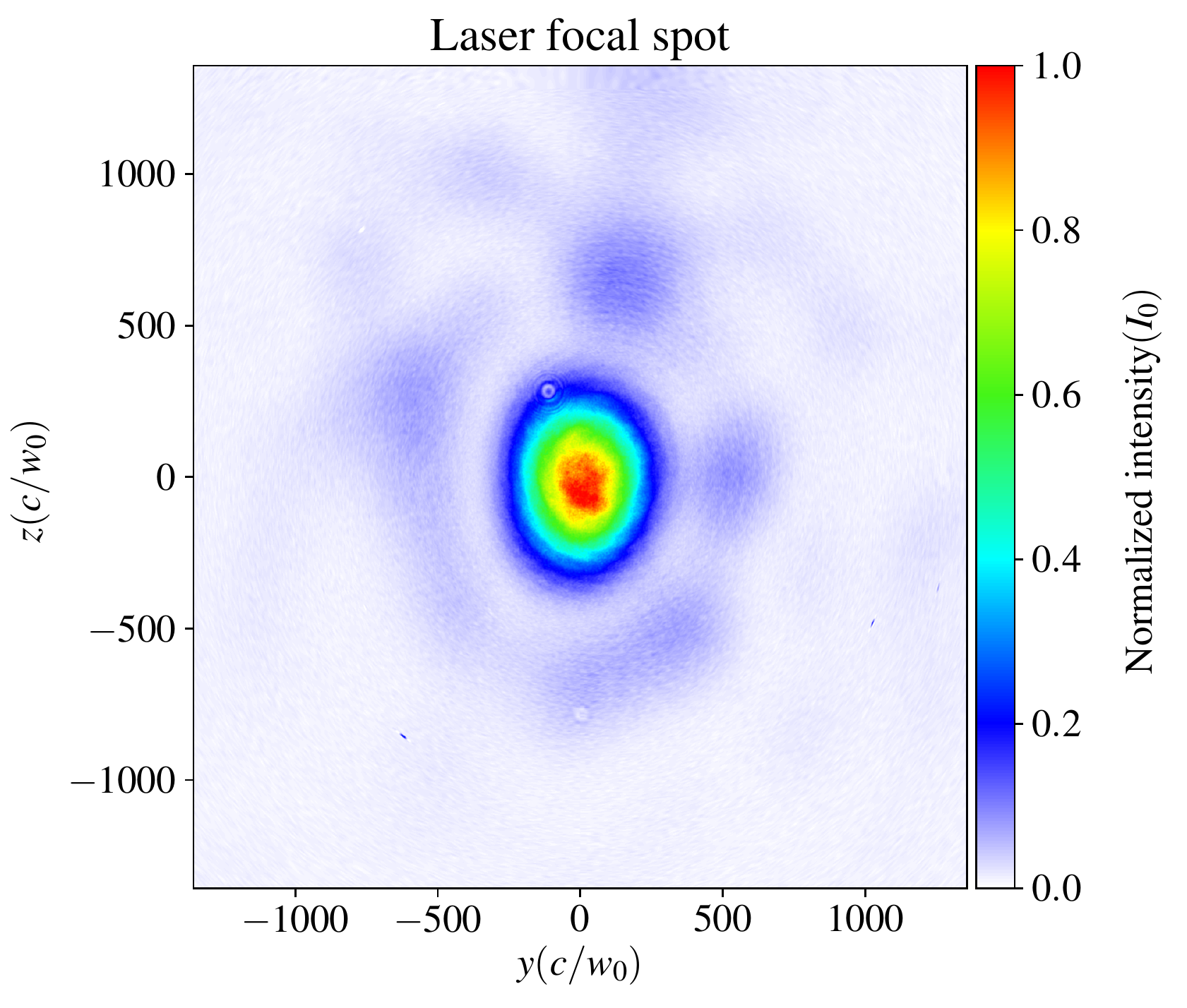}
\captionof{figure}{Laser  intensity  in the focal plan from experimental measurement.}
\label{laser}
%\end{figure}
\end{minipage}
%\hspace{0.6 cm}
\begin{minipage}{.45\textwidth}
\centering
%\begin{center}
\captionsetup{justification=centering}
\setlength{\tabcolsep}{10pt} % Default value: 6pt
\renewcommand{\arraystretch}{1.6}
\begin{tabular} {|l|l|}
  \hline
  \multicolumn{2}{|c|}{Laser parameters} \\
  \hline
   $a_0$ & 2.49 \\ 
    $\lambda_0 =2 \pi c / \omega_0$ & 0.8e-6 [m] \\ 
   Total energy  & 15 [J]  \\ 
    Laser duration & 30.e-15 [s] \\ 
    $I_0$  & 5.84e+19 [$W.cm^{-2}$] \\ 
    Waist & 40.e-6 [m] \\ 
  \hline
\end{tabular}
\vspace{0.5 cm}
  \caption{Laser parameters  with $\lambda_0$ the laser wavelength and $\omega_0$  its frequency. }

  \label{table}

%\end{center}
\end{minipage}
\end{table}

\begin{figure}[h!]
\centering
\includegraphics[width=\linewidth]{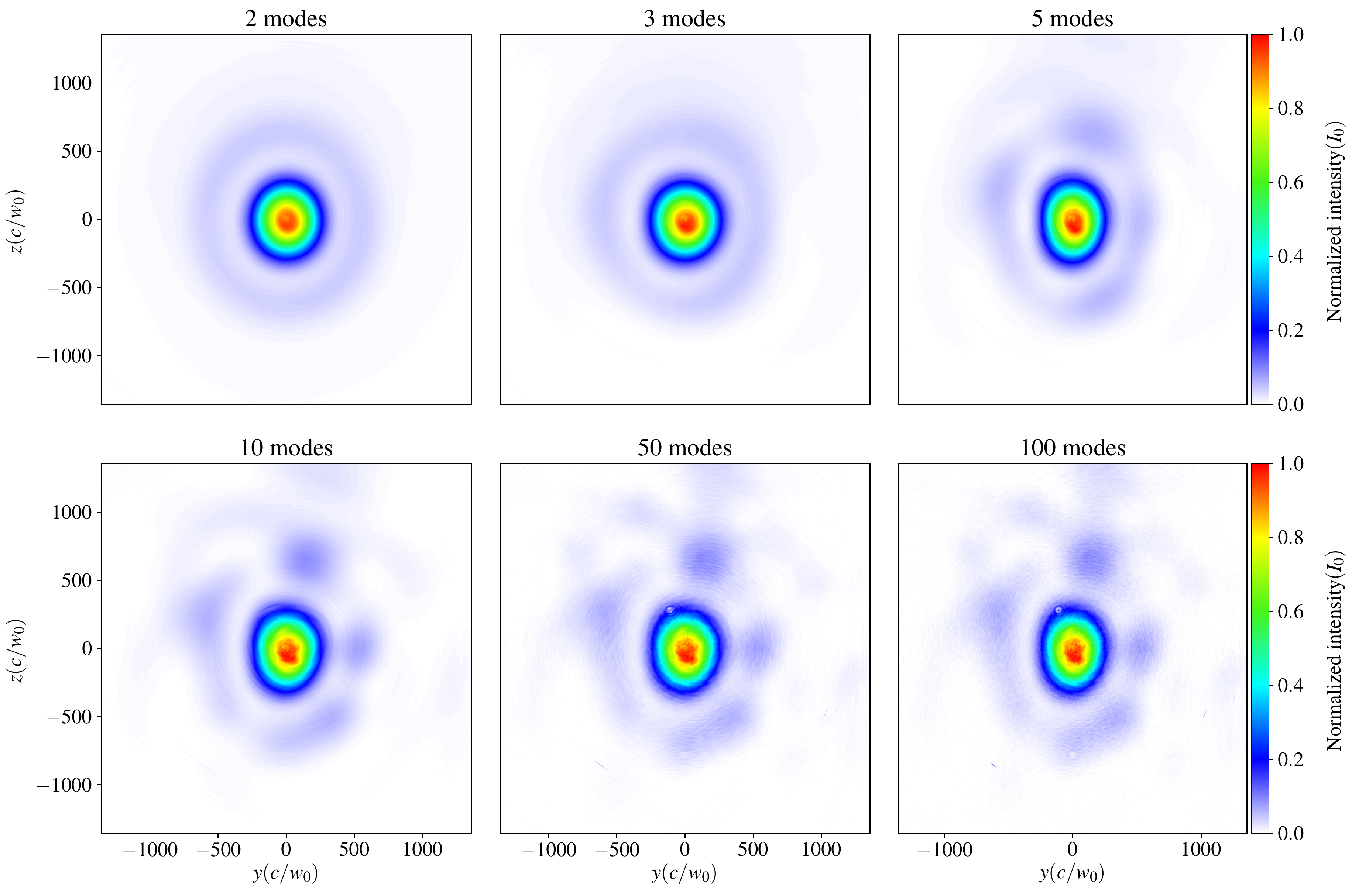}
\caption{Normalized intensity of reconstructed laser field  varying the number of modes .}
\label{allmodes}
\end{figure}

 Qualitatively, we can see from the reconstructed fields in figure \ref{allmodes} that the more modes we take into consideration, the closer we get to the real laser field and the more asymmetry  appears in the reconstructed field. For example, the 2 modes (mode 0 + mode 1) reconstructed field has an envelope that is perfectly axis-symmetric and that the intensity distribution surrounding the main spot is homogeneous and symmetric. However, more contrast and heterogeneity  appear in this area with higher number of modes. 
 
 \begin{figure}[h!]
\centering
\includegraphics[width= \linewidth]{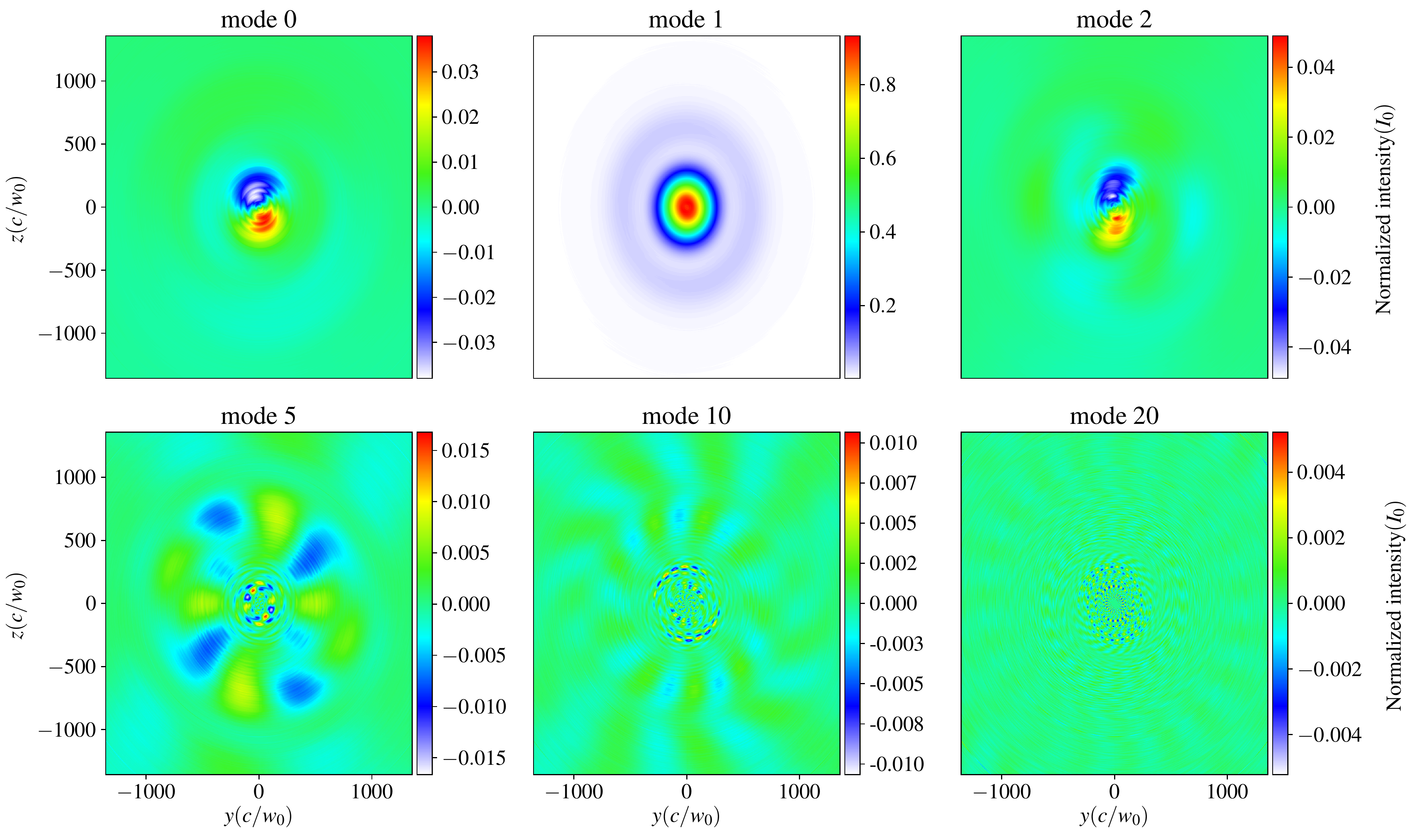}
\caption{ Normalized intensity of reconstructed laser fields in each mode separately.}
\label{seperate_modes}
\end{figure} 
 
 \begin{figure}[h!]
\centering
\includegraphics[width=0.5 \linewidth]{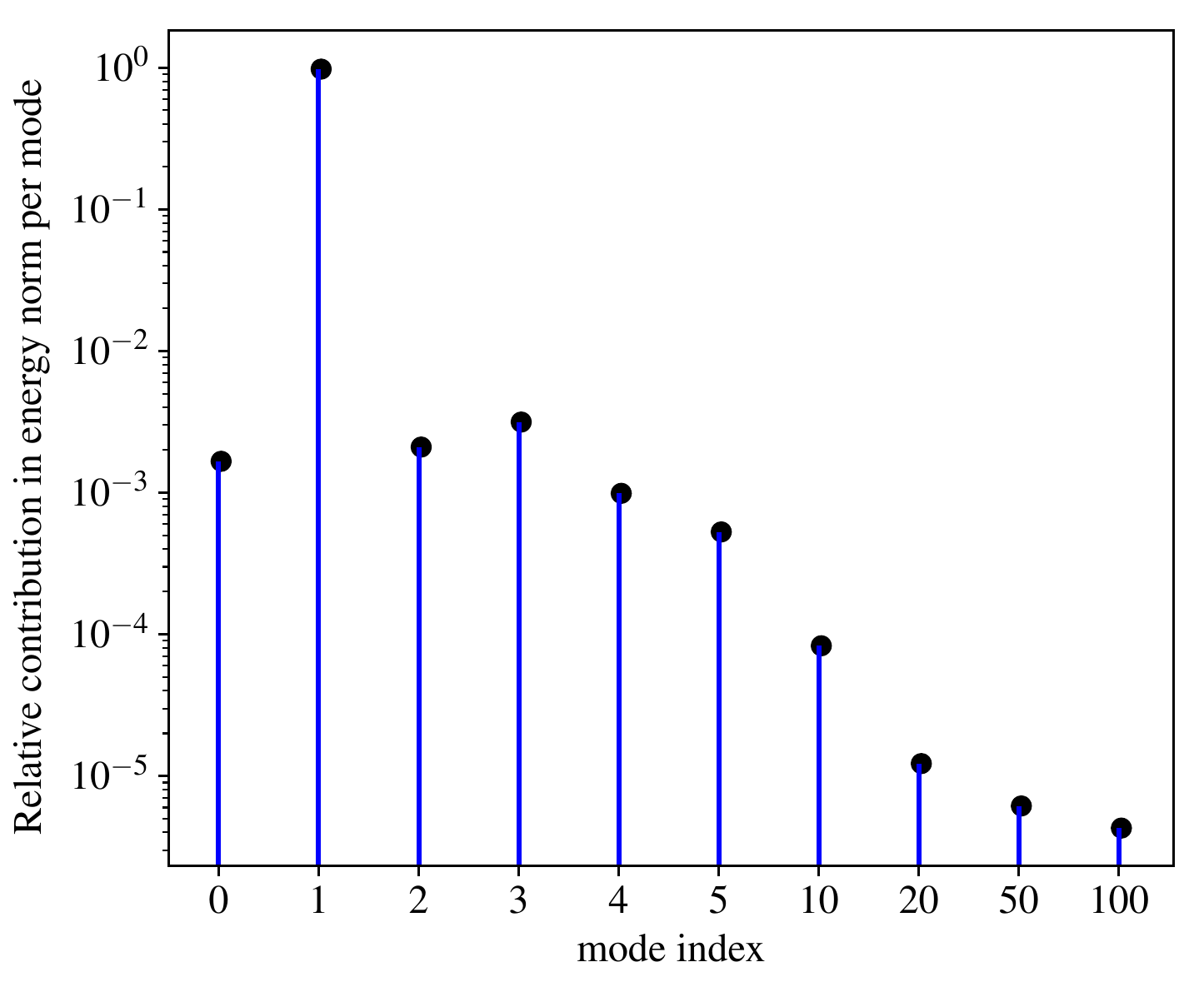}
\caption{Relative energy norm of the reconstructed fields for a sample of modes }
\label{distribution}
\end{figure} 

 Even though including higher modes in the reconstruction process is important to reproduce the heterogeneity in the intensity distribution, figure \ref{seperate_modes} shows that most of this intensity is represented exclusively by mode 1, which is the only non-zero mode in the case of cylindrically-symmetric laser, and other modes have lower contributions  (note the different color bars for each panel ). This is confirmed by the figure \ref{distribution}, that exhibits the distribution of the relative energy norm contained in each mode separately. The relative energy norm per mode was calculated as $\frac{||E_{mode}||_2}{||E_{laser}||_2}$
 % the  energy norm per mode divided by the energy norm of the laser from the measurement
  with the energy norm of a $2D$ discrete signal $x$  calculated according to: $||x||_2 = \sum_{i,j}  |x_{i,j}|^2$ . %
 The mode 1 has almost the entire energy of the laser and small portions are distributed unequally between the different modes: lower modes have more energy than the higher ones. Their contribution decreases with the number of modes.

 \begin{figure}[h!]
\centering
\captionsetup{justification=centering}
\includegraphics[width=0.4 \linewidth]{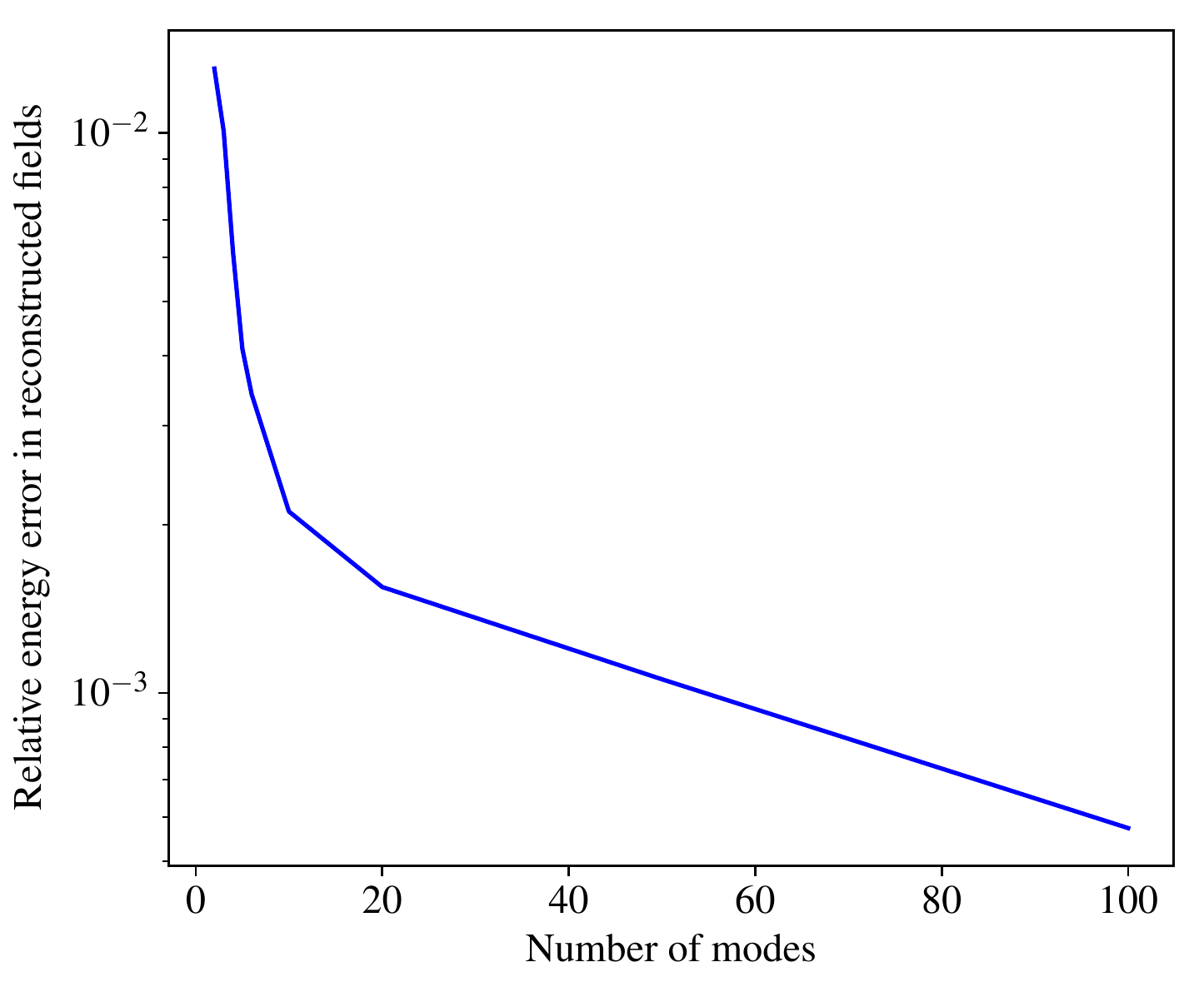}
\includegraphics[width=0.4 \linewidth]{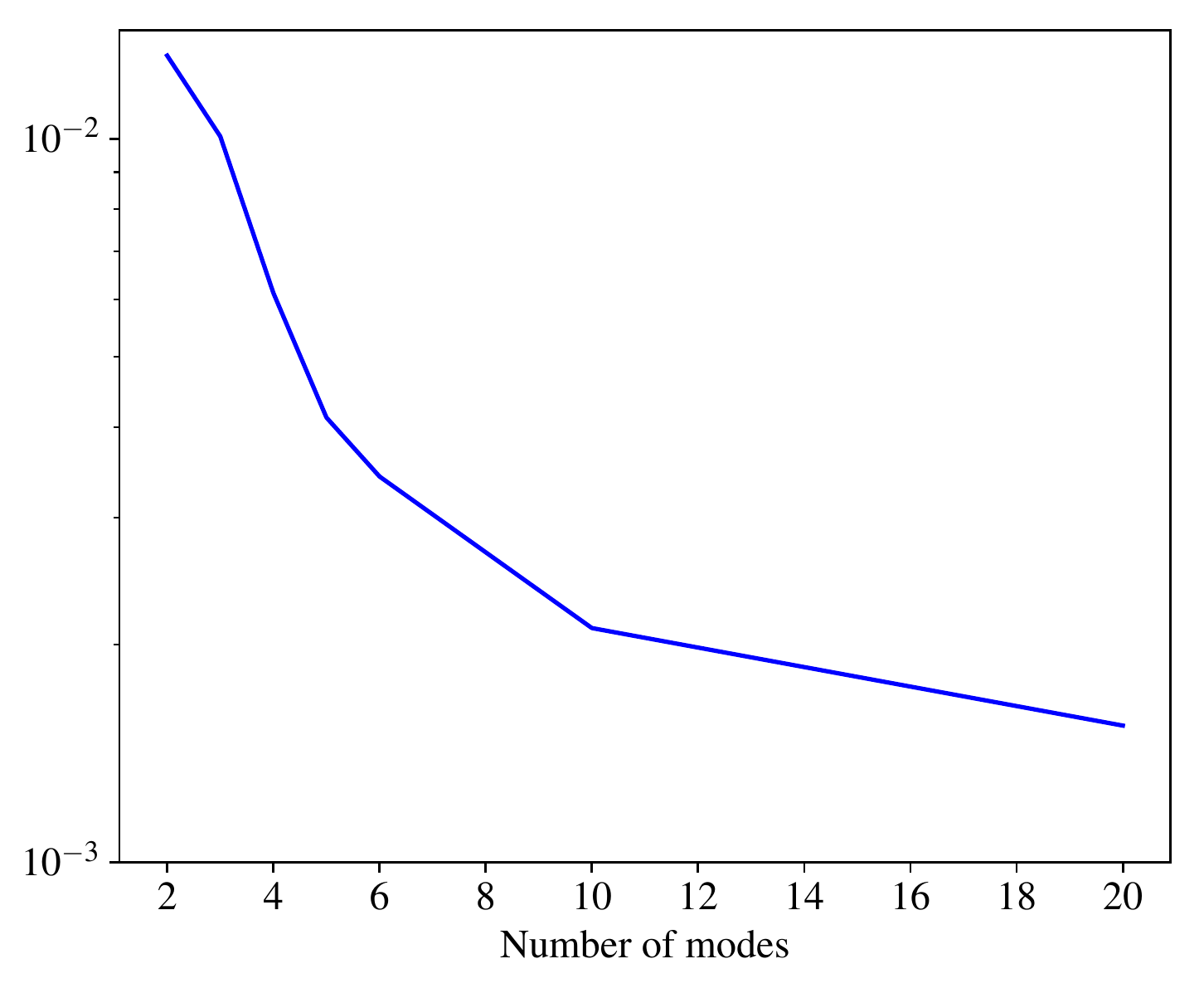}
\caption{Relative energy error of the reconstructed fields: on the left number of modes up to 100, on the right a close up to modes up to 20.}
\label{error}
\end{figure}

Figure \ref{error} shows the evolution of the relative energy norm error of the reconstructed field $\frac{||E_{reconstructed}||_2}{||E_{laser}||_2}$ in function of the number of modes used for the reconstruction with logarithmic scale along $y$ axis. Left panel shows that  the error decreases very quickly  up to the first 20 modes.  This means that the optimal  number of modes, which gives a good approximation of the real laser, is to be chosen  between 2 and 20. The panel on the right is  a close up of the panel on the left limiting the number of modes to the first 20 ones.  Even though, we restrain $m_{max}$ to 20, using 20 modes in the simulation is still very expensive for two reasons: First, the cost of the simulation with this geometry doesn't increase linearly with the number of modes because higher number of modes requires a higher number of particles per cell. Second, the Courant–Friedrichs–Lewy (CFL) condition of the azimuthal FDTD scheme depends on $m_{max}$, it becomes more strict with higher modes and implies smaller time step to avoid instabilities induced by the discretization scheme.   Therefore,  from figures \ref{allmodes} and \ref{error} it seems that 5 modes is a good compromise between precision and  a reasonable simulation cost. This analysis is a crucial step to the plugging of realistic lasers profiles in simulations run with the azimuthal cylindrical geometry

%\begin{figure}[h!]
%\centering
%\includegraphics[width=0.7\linewidth]{relative_B_enhanced.pdf}
%\caption{laser}
%\end{figure}
%
%
%\begin{figure}[h!]
%\centering
%\includegraphics[width=0.7\linewidth]{relative_B_resized_gridded.pdf}
%\caption{laser}
%\end{figure}
%
%
%\begin{figure}[h!]
%\centering
%\includegraphics[width=0.7\linewidth]{error.pdf}
%\caption{laser}
%\end{figure}
%
%
%
%\begin{figure}[h!]
%\centering
%\includegraphics[width=0.7\linewidth]{error_gridded.pdf}
%\caption{laser}
%\end{figure}

%\begin{figure}[h!]
%\centering
%\includegraphics[width=0.7\linewidth]{error_gridded_log.pdf}
%\caption{laser}
%\end{figure}

\section{Conclusion}
We have reviewed briefly the basics of the azimuthal Fourier field decomposition and how it can achieve a gain in numerical simulation cost with it. We have also proved its ability to describe realistic laser even with a low number of modes. Hence, this method combines precision, fidelity and speed up. However, a more accurate study should be carried not only on the focal spot of the laser but also with a correct reconstructed phase from the different measurements around the focal plan which may increase the needed number of modes to describe the laser evolution correctly.

%\section*{Acknowledgment}
\ack
We are grateful to Julien Derouillat for his dedication to help implementing this algorithm in \Smilei in an efficient way, to Frederic Perez for his help with the development of the post processing tool Happi to include this geometry, to Arnd Specka for his role of medium with the experimentalists  and for the fruitful discussions and to Dimitris  Papadopoulos and his team for providing us with the data. Imen Zemzemi is thankful to Haithem Kallala for the discussions about Fourier transform and Python tools to reconstruct the fields. 
F. Massimo was supported by P2IO LabEx (ANR-10-LABX-0038) in the framework “Investissements d’Avenir” (ANR-11-IDEX-0003-01) managed by the Agence Nationale de la Recherche (ANR, France).
%\begin{figure}[h!]
%\centering
%\includegraphics[width=0.7\linewidth]{error_log.pdf}
%\caption{laser}
%\end{figure}

\section*{References}

\bibliography{Bibliography}

\end{document}